\documentclass[12pt,aps,prd,showpacs,amsmath,amssymb]{revtex4}
\input epsf
\usepackage{graphicx}
\allowdisplaybreaks
\begin{document}
\title{Semileptonic Decays of $B^{**}_{s}$ mesons into $D_{s1}$ and $D^{*}_{s2}$ in HQET}
\author{Long-Fei Gan}
 \email{lfgan@nudt.edu.cn}
\author{Ming-Qiu Huang}
\affiliation{Department of Physics, College of Science, National University of Defense Technology, Changsha, Hunan 410073, People's Republic of China}
\date{\today}
\begin{abstract}
In the framework of the heavy quark effective theory, the leading order Isgur-Wise functions relevant to semileptonic decays of the orbitally $P$-wave excited $B_{s}$ meson states $B^{**}_{s}$, including the newly found narrow $B_{s1}(5830)$ and $B^{*}_{s2}(5840)$ states, into the ( $D_{s1}(2536)$, $D^{*}_{s2}(2573)$ ) doublet are calculated from QCD sum rules. With these universal form factors, the decay rates and branching ratios are also estimated.
\end{abstract}
 \maketitle

\section{Introduction}\label{sec1}
The investigation of the semileptonic $b\rightarrow c$ processes is an important source for the determination of the parameters of the standard model, such as Cabibbo-Kobayashi-Maskawa matrix element $|V_{cb}|$. They also provide valuable insight in quark dynamics in the nonperturbative domain of QCD. So a considerable amount of work has been carried out in the semileptonic decays of ground-state $B$ and $B_{s}$ mesons both experimentally and theoretically during the last two decades \cite{Neu92,LLSW97,Dea98,EFG99,Hua99,Col00,Hua04,AAO06,Gan09}. With the running energies of the colliders enhanced and the precision of the detectors improved, more and more excited $\bar{b}s$ mesons have been observed. In 2008, two orbitally excited narrow $\bar{b}s$ mesons, $B_{s1}(5830)$ and $B^{*}_{s2}(5840)$, were observed by the CDF and D0 Collaborations\cite{CDF08,D008}. This has inspired a lot of interest in their various decay modes\cite{ZZ08,LCL09,LCLZ09,Gan10}, among which the semileptonic decay modes are under our consideration.

The heavy quark effective theory (HQET) \cite{Neu94} has been proved to be an useful theoretical tool to describe the spectroscopy and matrix elements of mesons containing a heavy quark. In the heavy quark $m_{Q} \rightarrow \infty$ limit, the spin and parity of the heavy quark decouple from those of the light part. So the heavy-light mesons can be categorized according to the total angular momentum of the light degree of freedom $j_{q}$. Combining $j_{q}$ with the spin of the heavy quark yields a doublet of heavy-light meson states $j=j_{q}\pm1/2$. For $S$-wave ground states with $j_{q}=1/2$ ($L=0$), two states of negative parity form the $H$ doublet $(0^{-}, 1^{-})$. For $P$-wave excited states with $j_{q}=1/2$ or $j_{q}=3/2$ ($L=1$), four states of positive parity form the $S$ doublet $(0^{+}, 1^{+})$ and the $T$ doublet $(1^{+}, 2^{+})$. The four $P$-wave states of $\bar{b}s$ system are usually denoted as $B^{**}_{s}$ and the newly observed $B_{s1}(5830)$ and $B^{*}_{s2}(5840)$ mesons are considered as members of the $T$ doublet.

The QCD sum rule method \cite{Shi79} has been proved to be a successful nonperturbative approach which was widely used to investigate the properties of various types of hadrons \cite{WZ10,YT10,ZH10,WYCH11}. In our previous papers, we have employed it to investigate the semileptonic decays of ground-state $B$ mesons into some highly excited charmed meson doublets in the framework of HQET \cite{Gan09}, and then extended the discussion to the case of the semileptonic decays of $S$ and $T$ doublets of $\bar{b}s$ system into $H$ and $S$ doublets of the $\bar{c}s$ system \cite{Gan10}. We find that the semileptonic decay widths of a doublet of $\bar{b}s$ system into the same doublet in the $\bar{c}s$ system are significantly enhanced rather than those of its transitions into a different doublet. So it is reasonable to expect that the semileptonic decay widths of ($B_{s1}(5830)$, $B^{*}_{s2}(5840)$) into ($D_{s1}(2536)$,  $D^{*}_{s2}(2573)$) should be comparable with the previous decay modes. In order to verify this point, we investigate these semileptonic processes in this paper. As a byproduct, we also investigate the semileptonic decays of the other $B^{**}_{s}$ doublet into the same final states.

The remainder of this paper is organized as follows. After an introduction, we derive the formulas of the weak current matrix elements in HQET in Sec. \ref{sec2}. Then we deduce the three-point sum rules for the relevant universal form factors in Sec. \ref{sec3}. In Sec. \ref{sec4} we give the numerical results and discussions. The decay rates and branching ratios are also estimated in the final section.

\section{Parametrization of decay matrix elements in HQET} \label{sec2}
The semileptonic decay rate of a $B_{s}$ meson transition into a $D_{s}$ meson is determined by the corresponding matrix elements of the weak vector and axial-vector currents ($V^{\mu} = \overline{c}\gamma^{\mu}b$ and $A^{\mu} = \overline{c}\gamma^{\mu}\gamma_{5}b$) between them. These hadronic matrix elements can be parametrized in terms of form factors. In HQET, the classification of these form factors has been simplified greatly. For the transitions between
the $S$ doublet of the $\bar{b}s$ mesons and the $T$ doublet of $\bar{c}s$ mesons, the weak matrix elements can be parametrized in terms of one Isgur-Wise function at the leading order of the heavy quark expansion. But for the transitions between ($B_{s1}(5830)$, $B^{*}_{s2}(5840)$) and ($D_{s1}(2536)$, $D^{*}_{s2}(2573)$), we need two Isgur-Wise functions to parameterize the corresponding hadronic matrix elements at the leading order of the heavy quark expansion.

According to the formalism given in Ref. \cite{Fal92}, the heavy-light meson doublets can be expressed as effective operators. For the processes we considered, two heavy-light meson doublets $S$ and $T$ are involved. The operators $P_{0}$ and $P'_{1\mu}$ that annihilate members of the $S$ doublet with four-velocity $v$ are, in the form,
\begin{equation}\label{operator3}
 S_{v}=\frac{1+\rlap/v}{2}[P'_{1\mu}\gamma^{\mu}\gamma_{5}+P_{0}].
\end{equation}
The fields $P^{\nu}$ and $P^{*\mu\nu}$ that annihilate members of the $T$ doublet with four-velocity $v$ are in the representation
\begin{equation}\label{operator1}
 T^{\mu}_{v}=\frac{1+\rlap/v}{2}\{P^{*\mu\nu}\gamma_{\nu}-\sqrt{\frac{3}{2}}P^{\nu}\gamma_{5}
 [g^{\mu}_{\nu}-\frac{1}{3}\gamma_{\nu}(\gamma^{\mu}-v^{\mu})]\},
\end{equation}
where $\rlap/v = v\cdot\gamma$. At the leading order of the heavy quark expansion, the hadronic matrix elements of the weak currents between states in the doublets $S_{v}$ and $T_{v'}$ can be calculated from
\begin{equation}\label{trace1}
 \bar{h}^{(c)}_{v'}\Gamma h^{(b)}_{v} = \eta(y) \mathrm{Tr}\{v_{\sigma}
 \overline{T}^{(c)\sigma}_{v'}\Gamma S^{(b)}_{v}\},
\end{equation}
while the corresponding matrix elements between states annihilated by fields in $T_{v}$ and $T_{v'}$ are derived from
\begin{equation}\label{trace2}
 \bar{h}^{(c)}_{v'}\Gamma h^{(b)}_{v}=-\mathrm{Tr}\{(\xi_{1}(y)g_{\alpha\sigma}-\xi_{2}(y)v'_{\alpha}v_{\sigma})
 \overline{T}^{(c)\sigma}_{v'}\Gamma T^{(b)\alpha}_{v}\},
\end{equation}
where $h^{(Q)}_{v,v'}$ are the heavy quark fields in HQET, and $\overline{T}_{v'}=\gamma_{0}T_{v'}^{\dag}\gamma_{0}$. $v$ is the velocity of the initial meson and $v'$ is the velocity of the final meson in each process. The Isgur-Wise form factors $\eta(y)$, $\xi_{1}(y)$, and $\xi_{2}(y)$ are universal functions of the product of velocities $y(=v\cdot v')$. Here we should notice that each side of Eqs. (\ref{trace1}) and (\ref{trace2}) is understood to be inserted between the corresponding initial $B_{s}$ and final $D_{s}$ states. The hadronic matrix elements of $B_{s0}(B^{'}_{s1})\rightarrow D_{s1}(D^{*}_{s2})\ell\overline{\nu}$ can be derived directly from the trace formalism (\ref{trace1}) and are given as
\begin{align}
\label{matrix9}
\frac{\langle D_{s1}(v', \epsilon)|(V-A)^{\mu}|B_{s0}(v)\rangle}{\sqrt{m_{B_{s0}}m_{D_{s1}}}} = &
\sqrt{\frac{1}{6}}\eta(y)
\epsilon^{*}_{\beta}[(y^{2}-1)g^{\beta\mu}-(y+2)v'^{\mu}v^{\beta}+3v^{\mu}v^{\beta}\nonumber\\&+i(y-1)v_{\rho}v'_{\tau} \varepsilon^{\beta\mu\rho\tau}],
\\\label{matrix10}
\frac{\langle D^{*}_{s2}(v', \epsilon)|(V-A)^{\mu}|B_{s0}(v)\rangle}{\sqrt{m_{B_{s0}}m_{D^{*}_{s2}}}} = &
\eta(y)v^{\alpha}\epsilon^{*}_{\alpha\beta}[(1-y)g^{\beta\mu}+v'^{\mu}v^{\beta}-iv_{\rho}v'_{\tau}\varepsilon^ {\beta\mu\rho\tau}],
\\\label{matrix11}
\frac{\langle D_{s1}(v',\epsilon')|(V-A)^{\mu}|B'_{s1}(v,\epsilon)\rangle}{\sqrt{m_{B'_{s1}}m_{D_{s1}}}} = &
\sqrt{\frac{1}{6}}\eta(y)\epsilon'^{*}_{\beta}\epsilon_{\sigma}[-(y-1)(2g^{\sigma\mu}v^{\beta}+g^{\beta\mu}v'^{\sigma}) +3v'^{\sigma}v^{\mu}v^{\beta}\nonumber\\&+(y-1)(v^{\mu}+v'^{\mu})g^{\sigma\beta}+i(y+1)
\varepsilon^{\beta\sigma\mu\tau}(v_{\tau}-v'_{\tau})\nonumber\\&-2iv^{\mu}v_{\rho}v'_{\tau}\varepsilon ^{\beta\sigma\rho\tau}-iv^{\beta}v_{\rho}v'_{\tau}\varepsilon^{\sigma\mu\rho\tau}],
\\\label{matrix12}
\frac{\langle D^{*}_{s2}(v',\epsilon')|(V-A)^{\mu}|B'_{s1}(v,\epsilon)\rangle}{\sqrt{m_{B'_{s1}}m_{D^{*}_{s2}}}} = &
\eta(y)v^{\alpha}\epsilon'^{*}_{\alpha\beta}\epsilon_{\sigma}[g^{\beta\mu}v'^{\sigma}-
g^{\sigma\mu}v^{\beta}+(v^{\mu}-v'^{\mu})g^{\beta\sigma}\nonumber\\&-i\varepsilon^{\beta\sigma\mu\tau}(v_{\tau} -v'_{\tau})].
\end{align}
The hadronic matrix elements of $B_{s1}\rightarrow D_{s1}(D^{*}_{s2})\ell\overline{\nu}$ are calculated similarly from Eq. (\ref{trace2}) as follows:
\begin{align}\label{matrix1}
\frac{\langle D_{s1}(v',\epsilon')|V^{\mu}|B_{s1}(v,\epsilon)\rangle}{\sqrt{m_{B_{s1}}m_{D_{s1}}}} =&
\frac{1}{6}\epsilon'_{\rho}\epsilon_{\beta}\{\xi_{1}(y)[(2y+1)g^{\mu\rho}v'^{\beta}+(y+5)g^{\beta\rho}
(v'^{\mu}+v^{\mu})\nonumber\\&+(2y+1)v^{\rho}g^{\beta\mu}-3v^{\rho}v'^{\beta}(v'^{\mu}+v^{\mu})]
\nonumber\\&+\xi_{2}(y)[-2(y^{2}-1)g^{\mu\rho}v'^{\beta}+(1-y^{2})g^{\beta\rho}(v'^{\mu}+v^{\mu})\nonumber\\&-2(y^{2}-1)
v^{\rho}g^{\beta\mu}+3(y-2)v^{\rho}v'^{\beta}(v'^{\mu}+v^{\mu})]\},
\\\label{matrix2}
\frac{\langle D_{s1}(v',\epsilon')|A^{\mu}|B_{s1}(v,\epsilon)\rangle}{\sqrt{m_{B_{s1}}m_{D_{s1}}}} =&
\frac{i}{6}\epsilon'_{\rho}\epsilon_{\beta}\{\xi_{1}(y)[-v'^{\beta}\varepsilon^{\mu\rho\alpha\sigma}v_{\alpha}
v'_{\sigma}+2(v^{\mu}-v'^{\mu})\varepsilon^{\beta\rho\alpha\sigma}v_{\alpha}v'_{\sigma}\nonumber\\&+v^{\rho}
\varepsilon^{\beta\mu\alpha\sigma}v_{\alpha}v'_{\sigma}-(y+2)\varepsilon^{\beta\mu\rho\sigma}(v'_{\sigma}
+v_{\sigma})]\nonumber\\&-\xi_{2}(y)[(y^{2}-1)\varepsilon^{\beta\mu\rho\sigma}(v'_{\sigma}+v_{\sigma})+(y+1)v'^{\beta}
\varepsilon^{\mu\rho\alpha\sigma}v_{\alpha}v'_{\sigma}\nonumber\\&-2(y+1)(v^{\mu}-v'^{\mu})
\varepsilon^{\beta\rho\alpha\sigma}v_{\alpha}v'_{\sigma}-(y+1)v^{\rho}\varepsilon^{\beta\mu\alpha\sigma}
v_{\alpha}v'_{\sigma}]\},
\\\label{matrix3}
\frac{\langle D^{*}_{s2}(v',\epsilon')|V^{\mu}|B_{s1}(v,\epsilon)\rangle}{\sqrt{m_{B_{s1}}m_{D^{*}_{s2}}}} =&
\frac{i}{\sqrt{6}}\epsilon'_{\sigma\rho}\epsilon_{\beta}\{\xi_{1}(y)[g^{\beta\mu}\varepsilon^{\rho\sigma\alpha\tau} v_{\alpha} v'_{\tau}-g^{\beta\rho}\varepsilon^{\mu\sigma\alpha\tau}v_{\alpha}v'_{\tau}
+2g^{\beta\sigma}\varepsilon^{\mu\rho\alpha\tau}v_{\alpha}v'_{\tau}\nonumber\\&+g^{\mu\rho}
\varepsilon^{\beta\sigma\alpha\tau}v_{\alpha}v'_{\tau}+g^{\mu\sigma}\varepsilon^{\beta\rho\alpha\tau}
v_{\alpha}v'_{\tau}+v'^{\beta}\varepsilon^{\mu\rho\sigma\alpha}v_{\alpha}-v^{\mu}
\varepsilon^{\beta\rho\sigma\alpha}v'_{\alpha}\nonumber\\&-v'^{\mu}\varepsilon^{\beta\rho\sigma\alpha}v_{\alpha}+
v^{\rho}\varepsilon^{\beta\mu\sigma\alpha}v'_{\alpha}+v^{\sigma}\varepsilon^{\beta\mu\rho\alpha}v_{\alpha}
+(y-1)\varepsilon^{\beta\mu\rho\sigma}]\nonumber\\&
+\xi_{2}(y)[v^{\sigma}\varepsilon^{\beta\mu\rho\alpha}v'_{\alpha}-v^{\sigma}v'^{\beta}
\varepsilon^{\mu\rho\alpha\tau}v_{\alpha}v'_{\tau}\nonumber\\&-2v^{\sigma}v'^{\mu}
\varepsilon^{\beta\rho\alpha\tau}v_{\alpha}v'_{\tau}-yv^{\sigma}\varepsilon^{\beta\mu\rho\alpha}
v'_{\alpha}-(y-1)v^{\sigma}\varepsilon^{\beta\mu\rho\alpha}v_{\alpha}]\},
\\\label{matrix4}
\frac{\langle D^{*}_{s2}(v',\epsilon')|A^{\mu}|B_{s1}(v,\epsilon)\rangle}{\sqrt{m_{B_{s1}}m_{D^{*}_{s2}}}} =&
\frac{1}{\sqrt{6}}\epsilon'_{\sigma\rho}\epsilon_{\beta}\{\xi_{1}(y)[-v^{\rho}g^{\mu\sigma}v'^{\beta}
+2g^{\beta\sigma}((y+1)g^{\mu\rho}-v^{\rho}v'^{\mu})\nonumber\\&-v^{\mu}v^{\sigma}g^{\beta\rho}
+v^{\rho}v^{\sigma}g^{\beta\mu}+(y+1)g^{\beta\rho}g^{\mu\sigma}]
\nonumber\\&+\xi_{2}(y)[-2(1+y)v^{\sigma}g^{\mu\rho}v'^{\beta}-(y+1)v^{\sigma}g^{\beta\rho}v'^{\mu}
\nonumber\\&-(y+1)v^{\rho}v^{\sigma}g^{\beta\mu}+(y+1)v^{\mu}v^{\sigma}g^{\beta\rho}
+3v^{\rho}v^{\sigma}v'^{\beta}v'^{\mu}]\}.
\end{align}
For the decays $B^{*}_{s2}\rightarrow D_{s1}(D^{*}_{s2})\ell\overline{\nu}$, the corresponding hadronic matrix elements are as follows:
\begin{align}
\label{matrix5}
\frac{\langle D_{s1}(v',\epsilon')|V^{\mu}|B^{*}_{s2}(v,\epsilon)\rangle}{\sqrt{m_{B^{*}_{s2}}m_{D_{s1}}}} =&
-\frac{i}{\sqrt{6}}\epsilon'_{\rho}\epsilon_{\alpha\beta}\{\xi_{1}(y)[g^{\alpha\mu}
\varepsilon^{\beta\rho\sigma\tau}v_{\sigma}
v'_{\tau}+2g^{\alpha\rho}\varepsilon^{\beta\mu\sigma\tau}v_{\sigma}v'_{\tau}
+g^{\beta\mu}\varepsilon^{\alpha\rho\sigma\tau}v_{\sigma}v'_{\tau}\nonumber\\&-g^{\beta\rho}
\varepsilon^{\alpha\mu\sigma\tau}v_{\sigma}v'_{\tau}+g^{\mu\rho}\varepsilon^{\alpha\beta\sigma\tau}
v_{\sigma}v'_{\tau}-v'^{\alpha}\varepsilon^{\beta\mu\rho\tau}v'_{\tau}-v'^{\beta}
\varepsilon^{\alpha\mu\rho\tau}v_{\tau}\nonumber\\&+v^{\mu}\varepsilon^{\alpha\beta\rho\tau}v'_{\tau}+
v'^{\mu}\varepsilon^{\alpha\beta\rho\tau}v_{\tau}-v^{\rho}\varepsilon^{\alpha\beta\mu\tau}v'_{\tau}
+(y-1)\varepsilon^{\alpha\beta\mu\rho}]\nonumber\\&
+\xi_{2}(y)[(y-1)v'^{\alpha}\varepsilon^{\beta\mu\rho\tau}v_{\tau}+(y-1)v'^{\alpha}
\varepsilon^{\beta\mu\rho\tau}v'_{\tau}\nonumber\\&-2v^{\mu}v'^{\alpha}
\varepsilon^{\beta\rho\sigma\tau}v_{\sigma}v'_{\tau}-v^{\rho}v'^{\alpha}\varepsilon^{\beta\mu\sigma\tau}
v_{\sigma}v'^{\tau}]\},
\\\label{matrix6}
\frac{\langle D_{s1}(v',\epsilon')|A^{\mu}|B^{*}_{s2}(v,\epsilon)\rangle}{\sqrt{m_{B^{*}_{s2}}m_{D_{s1}}}} =&
\frac{1}{\sqrt{6}}\epsilon'_{\rho}\epsilon_{\alpha\beta}\{\xi_{1}(y)[g^{\mu\rho}v'^{\alpha}v'^{\beta}
-g^{\beta\rho}v'^{\alpha}v'^{\mu}+2g^{\alpha\rho}((y+1)g^{\beta\mu}-v^{\mu}v'^{\beta})\nonumber\\&
+g^{\alpha\mu}((y+1)g^{\beta\rho}-v^{\rho}v'^{\beta})]
\nonumber\\&+\xi_{2}(y)[-(1+y)g^{\mu\rho}v'^{\alpha}v'^{\beta}-(y+1)g^{\beta\rho}v'^{\alpha}(v^{\mu}-
v'^{\mu})\nonumber\\&+2(y+1)g^{\beta\mu}v^{\rho}v'^{\alpha}
+3v^{\mu}v^{\rho}v'^{\alpha}v'^{\beta}]\},
\\\label{matrix7}
\frac{\langle D^{*}_{s2}(v',\epsilon')|V^{\mu}|B^{*}_{s2}(v,\epsilon)\rangle}{\sqrt{m_{B^{*}_{s2}}m_{D^{*}_{s2}}}} = &
\epsilon'_{\sigma\rho}\epsilon_{\alpha\beta}\{\xi_{1}(y)g^{\alpha\sigma}[g^{\mu\rho}v'^{\beta}
-g^{\beta\rho}(v'^{\mu}+v^{\mu})+v^{\rho}g^{\beta\mu}]
\nonumber\\&-\xi_{2}(y)v^{\sigma}v'^{\alpha}[g^{\mu\rho}v'^{\beta}-g^{\beta\rho}(v'^{\mu}
+v^{\mu})+v^{\rho}g^{\beta\mu}]\},
\\\label{matrix8}
\frac{\langle D^{*}_{s2}(v',\epsilon')|A^{\mu}|B^{*}_{s2}(v,\epsilon)\rangle}{\sqrt{m_{B^{*}_{s2}}m_{D^{*}_{s2}}}} = & i \epsilon'_{\sigma\rho}\epsilon_{\alpha\beta}\{\xi_{1}(y)g^{\alpha\sigma}\varepsilon^{\beta\mu\rho\tau}
(v'_{\tau}+v_{\tau})-\xi_{2}(y)v^{\sigma}v'^{\alpha}\varepsilon^{\beta\mu\rho\tau}
(v'_{\tau}+v_{\tau})\}
\end{align}
In these matrix elements, $\epsilon_{\alpha}$ ($\epsilon'_{\alpha}$) is the polarization vector of the initial (final) vector meson and $\epsilon_{\alpha\beta}$ ($\epsilon'_{\sigma\rho}$) is the polarization tensor of the initial (final) tensor meson. The only unknown factors in the matrix elements above are $\eta(y)$, $\xi_{1}(y)$ and $\xi_{2}(y)$ which should be determined by nonperturbative methods. In the following section, we will apply the QCD sum rule approach to estimate them.

\section{Form factors from HQET sum rules}\label{sec3}
In order to apply QCD sum rules to study these heavy mesons, we must choose appropriate interpolating currents to represent these states. Here we adopt the interpolating currents proposed in Ref. \cite{Dai97} based on the study of Bethe-Salpeter equations for heavy mesons in HQET. Following the remarks given in Ref. \cite{Gan09}, we take the interpolating currents that create heavy mesons in the $T$ doublet and $S$ doublet as
\begin{align}
\label{current1}
J^{\dag\alpha}_{1,+,3/2} &=(-i)\sqrt{\frac{3}{4}}\bar{h}_{v}\gamma_{5}(D^{\alpha}_{t}-\frac{1}{3} \gamma^{\alpha}_{t}\!\not\!\!{D}_{t})s,
\\ \label{current2}
J^{\dag\alpha\beta}_{2,+,3/2} &=\frac{(-i)}{\sqrt{2}}T^{\alpha\beta,\mu\nu}\bar{h}_{v}\gamma_{t\mu}D_{t\nu}s,
\\ \label{current3}
J^{\dag}_{0,+,1/2} &=\frac{1}{\sqrt{2}}\bar{h}_{v}(-i)\!\not\!\!{D}_{t}s,
\\ \label{current4}
J^{\dag\alpha}_{1,+,1/2} &=\frac{1}{\sqrt{2}}\bar{h}_{v}\gamma_{5}\gamma^{\alpha}_{t}(-i)\!\not\!\!{D}_{t}s,
\end{align}
where $D^{\alpha}_{t}=D^{\alpha}-v^{\alpha}(v\cdot D)$ is the transverse component of the covariant derivative with respect to the velocity of the meson. The tensor $T^{\alpha\beta,\mu\nu}$ is used to symmetrize the indices and is given by
\begin{equation}\label{tensor1}
T^{\alpha\beta,\mu\nu}=\frac{1}{2}(g^{\alpha\mu}_{t}g^{\beta\nu}_{t}
+g^{\alpha\nu}_{t}g^{\beta\mu}_{t})-\frac{1}{3}g^{\alpha\beta}_{t}g^{\mu\nu}_{t},
\end{equation}
where $g^{\alpha\beta}_{t} = g^{\alpha\beta} - v^{\alpha} v^{\beta}$ is the transverse part of the metric tensor relative to the velocity of the heavy meson.

These currents have non-vanishing projections only to the corresponding states of the HQET in the $m_{Q}\rightarrow\infty$ limit, without mixing with states of the same quantum number but different $s_{l}$. Thus we can define one-particle-current couplings as follows:
\begin{align}
\label{const1}
\langle H_{s1}(v,\varepsilon)|J^{\alpha}_{1,+,3/2}|0\rangle &= f_{1,+,3/2}\sqrt{m_{H_{s1}}} \varepsilon^{*\alpha}, &\text{ for}\; J^{P}=1^{+};
\\ \label{const2}
\langle H^{*}_{s2}(v,\varepsilon)|J^{\alpha\beta}_{2,+,3/2}|0\rangle &= f_{2,+,3/2} \sqrt{m_{H^{*}_{s2}}} \varepsilon^{*\alpha\beta}, &\text{ for}\; J^{P}=2^{+};
\\ \label{const3}
\langle H_{s1}(v,\varepsilon)|J^{\alpha}_{0,+,1/2}|0\rangle &= f_{0,+,1/2}\sqrt{m_{H_{s0}}}, &\text{ for}\; J^{P}=0^{+};
\\ \label{const4}
\langle H^{*}_{s2}(v,\varepsilon)|J^{\alpha}_{1,+,1/2}|0\rangle &= f_{1,+,1/2}\sqrt{m_{H^{*}_{s2}}} \varepsilon^{*\alpha}, &\text{ for}\; J^{P}=1^{+}.
\end{align}
The decay constants $f_{1,+,3/2}$, $f_{2,+,3/2}$, $f_{0,+,1/2}$, and $f_{1,+,1/2}$ are low-energy parameters which are determined by the dynamics of the light degree of freedom.

With these currents, we can now estimate the Isgur-Wise functions $\xi_{1}(y)$, $\xi_{2}(y)$, and $\eta(y)$ from QCD sum rules. Let us consider $\xi_{1}(y)$ and $\xi_{2}(y)$ first. The jumping-off point is the following three-point correlation function:
\begin{align}\label{correlator1}
\Xi^{\sigma\mu\alpha}(\omega,\omega^{'},y) & =i^{2}\int d^{4}xd^{4}ze^{i(k^{'}\cdot x-k\cdot z)}\langle0|T[J^{\sigma}_{1,+,3/2}(x) J^{\mu(v,v^{'})}_{V,A}(0)J^{\alpha\dag}_{1,+,3/2}(z)|0\rangle\nonumber\\ & = \Xi_{1}(\omega,\omega^{'},y)\mathcal {L}^{\sigma\mu\alpha}_{\xi_{1}(V,A)}+\Xi_{2}(\omega,\omega^{'},y)\mathcal{L}^{\sigma\mu\alpha}_{\xi_{2}(V,A)},
\end{align}
where $J^{\mu(v,v^{'})}_{V}=h(v^{'})\gamma^{\mu}h(v)$ and $J^{\mu(v,v^{'})}_{A}=h(v^{'})\gamma^{\mu}\gamma_{5}h(v)$ are the weak currents, $J^{\sigma(\alpha)}_{1,+,3/2}$ is the interpolating current defined in Eq. (\ref{current1}). $\Xi_{i}(\omega,\omega^{'},y)$ ($i=1, 2$) are analytic functions in $\omega=2v \cdot k$ and $\omega'=2v' \cdot k'$, and are not continual when $\omega$ and $\omega'$ locate on the positive real axis. $k$($=P-m_{b}v$) and $k^{'}$($=P'-m_{c}v'$) are the residual momenta of the initial and final meson states, respectively. The scalar functions $\Xi_{i}(\omega,\omega^{'},y)$ ($i=1, 2$) also depend on the velocity transfer $y=v \cdot v'$. $\mathcal {L}^{\sigma\mu\alpha}_{\xi_{1}(V,A)}$ and $\mathcal {L}^{\sigma\mu\alpha}_{\xi_{2}(V,A)}$ are Lorentz structures.

To calculate the phenomenological or the physical part of the correlator (\ref{correlator1}), we insert two complete sets of intermediate states with the same quantum number as the current $J_{1,+,3/2}(x)$ and isolate the contribution from the double pole at $\omega=2 \bar{\Lambda}$, $\omega'=2 \bar{\Lambda}$:
\begin{align}\label{pheno}
\Xi^{\sigma\mu\alpha}(\omega,\omega^{'},y) & = \frac{f^{2}_{1,+,3/2}}{(2\bar{\Lambda}-\omega-i \epsilon)(2\bar{\Lambda}-\omega'-i \epsilon)}[\xi_{1}(y)\mathcal
{L}^{\sigma\mu\alpha}_{\xi_{1}}+\xi_{2}(y)\mathcal{L}^{\sigma\mu\alpha}_{\xi_{2}}]+\cdots,
\end{align}
where ``$\cdots$" denotes contributions from higher resonances and continuum states, $f_{1,+,3/2}$ is the decay constant defined in Eq. (\ref{const1}). As we can see from the equations (\ref{correlator1}) and (\ref{pheno}), the pole contribution to $\Xi_{1}(\omega,\omega^{'},y)$ is proportional to the universal function $\xi_{1}(y)$ while that to $\Xi_{2}(\omega,\omega^{'},y)$ is proportional to $\xi_{2}(y)$. By isolating the different Lorentz structures, the sum rules for $\xi_{1}(y)$ and $\xi_{2}(y)$ can be constructed from $\Xi_{1}(\omega,\omega^{'},y)$ and $\Xi_{2}(\omega,\omega^{'},y)$, respectively.

The theoretical side of the correlator is calculated by means of the operator product expansion. The perturbative part are usually expressed as a double dispersion integral in $\nu$ and $\nu^{'}$ plus possible subtraction terms. Therefore the theoretical expressions for the correlation functions in (\ref{correlator1}) are of the form
\begin{equation}\label{theo}
\Xi_{1,2}^{\text{theo}}(\omega,\omega^{'},y)\simeq\int d\nu d\nu^{'} \frac{\rho^{\text{pert}}_{1,2} (\nu,\nu^{'},y)} {(\nu-\omega-i\varepsilon) (\nu^{'}-\omega^{'}-i\varepsilon)}+ \text{subtractions} + \Xi^{\text{cond}}_{1,2}(\omega,\omega^{'},y).
\end{equation}
The perturbative spectral densities can be calculated straightforward from HQET Feynman rules. We consider only the leading order of perturbation here and the perturbative spectral densities of the two sum rules for
$\xi_{1}(y)$ and $\xi_{2}(y)$ are,
\begin{align}\label{perturb1}
\rho^{\text{pert}}_{\xi_{1}}(\nu,\nu',y) = &-\frac{3}{2^{8}\pi^{2}}\frac{1}{(y+1)^{5/2} (y-1)^{3/2}}[(\nu+\nu')-16m_{s}(y+1)](\nu^{2}-2y\nu\nu'+\nu'^{2})
\nonumber\\&\times\Theta(\nu)\Theta(\nu')\Theta(2y\nu\nu'-\nu^{2}-\nu'^{2}),
\end{align}
and
\begin{align}\label{perturb2}
\rho^{\text{pert}}_{\xi_{2}}(\nu,\nu^{'},y)= &-\frac{3}{2^{8}\pi^{2}}\frac{1}{(y+1)^{7/2}(y-1)^{5/2}} \{(\nu+\nu') [(4y-1)(\nu^{2}+\nu'^{2})-2(3y^{2}-y \nonumber\\ &+1)\nu\nu']-64m_{s}(y+1)[3y\nu^{2}-2\nu\nu'(2y^{2}+1) +2y\nu'^{2}]\}\nonumber\\&\times\Theta(\nu)\Theta(\nu')\Theta(2y\nu\nu'-\nu^{2}-\nu'^{2}),
\end{align}
respectively. Assuming quark-hadron duality, the contributions from higher resonances are usually approximated by the integrations of the perturbative spectral densities above some threshold. Equating the phenomenological and theoretical representations, the contributions of higher resonances in the phenomenological expression (\ref{pheno}) can be eliminated. Following the arguments in Refs. \cite{Neu92,Blo93}, we can not directly assume local duality between the perturbative and the hadronic spectral densities, but first integrate the spectral densities over the ``off-diagonal" variable $\nu_{-}=\nu-\nu^{'}$, keeping the ``diagonal" variable $\nu_{+}=\frac{\nu+\nu^{'}}{2}$ fixed. Then the quark-hadron duality is assumed for the integrations of the spectral densities in $\nu_{+}$. The integration region is restricted by the $\Theta$ functions above in terms of the variables $\nu_{-}$ and $\nu_{+}$ and usually the triangular region defined by the bounds: $0\leq \nu_{+}\leq \omega_{c}$, $-2\sqrt{\frac{y-1}{y+1}}\nu_{+}\leq \nu_{-}\leq
2\sqrt{\frac{y-1}{y+1}}\nu_{+}$ is chosen. A double Borel transformation in $\omega$ and $\omega^{'}$ is performed on both sides of the sum rules, in which for simplicity we take the Borel parameters equal
\cite{Neu92,Hua99,Col00}: $T_{1}=T_{2}=2T$. It eliminates the subtraction terms in the dispersion integral (\ref{theo}) and improves the convergence of the operator product expansion series. Our calculations are confined at the leading order of perturbation. Among the operators in the operator product expansion series, only those with dimension $D \leq  5$ are included. For the condensates of higher dimension ($D > 5$), their values are negligibly small and their contributions are suppressed by the double Borel transformation. So they can be safely omitted. Finally, we obtain the sum rules for the form factors $\xi_{1}(y)$ and $\xi_{2}(y)$ as follows:
\begin{align}
\label{rule1}
\xi_{1}(y)f^{2}_{1,+,3/2}e^{-(2\bar{\Lambda}_{1,+,3/2})/T}= &
\frac{1}{8\pi^{2}}\frac{1}{(y+1)^{3}}\int^{\omega_{c}}_{2m_{s}}d\nu_{+}e^{-\frac{\nu_{+}}{T}}
[\nu^{4}_{+}+m_{s}(y+1)\nu^{3}_{+}] \nonumber\\& -\frac{\langle g_{s} \bar{s}\sigma\cdot G s\rangle}{12}(1+\frac{m_{s}}{4T})-\frac{T}{32}\frac{3y+1}{(y+1)^{2}}
\langle\frac{\alpha_{s}}{\pi}GG\rangle,
\\ \label{rule2}
\xi_{2}(y)f^{2}_{1,+,3/2}e^{-(2\bar{\Lambda}_{1,+,3/2})/T}= &
\frac{1}{8\pi^{2}}\frac{1}{(y+1)^{4}}\int^{\omega_{c}}_{2m_{s}}d\nu_{+}e^{-\frac{\nu_{+}}{T}}
[3\nu^{4}_{+}+2m_{s}(y+1)\nu^{3}_{+}] \nonumber\\& -\frac{T}{24}\frac{2y+1}{(y+1)^{3}}
\langle\frac{\alpha_{s}}{\pi}GG\rangle,
\end{align}
where $\langle g_{s} \bar{s}\sigma\cdot G s\rangle=m^{2}_{0}\langle\bar{s}s\rangle$ with $\label{mcond}m^{2}_{0}=0.8 \mbox{GeV}^{2}$.

The derivation of the sum rule for the Isgur-Wise function $\eta(y)$ is completely similar. Only now the correlation function we need to consider is
\begin{equation}\label{correlator2}
i^{2}\int d^{4}xd^{4}ze^{i(k^{'}\cdot x-k\cdot z)}\langle0|T[J^{\alpha}_{1,+,3/2}(x) J^{\mu(v,v^{'})}_{V, A}(0) J^{\dag}_{0,+,1/2}(z)|0\rangle=\Xi(\omega,\omega^{'},y)\mathcal{L}^{\alpha\mu}_{V,A},
\end{equation}
where $J^{\mu(v,v^{'})}_{V, A}$ are also the weak currents, $J^{\alpha}_{1,+,3/2}$ and $J_{0,+,1/2}$ are the interpolating current defined in Eqs. (\ref{current1}) and (\ref{current3}). Here only one Lorentz structure $\mathcal{L}^{\alpha\mu}_{V,A}$ appears. By repeating the procedure above, we reach the sum rule for $\eta(y)$ as below:
\begin{align}\label{rule3}
\eta(y)f_{0,+,1/2}f_{1,+,3/2}e^{-(\bar{\Lambda}_{0,+,1/2}+\bar{\Lambda}_{1,+,3/2})/T}= &
\frac{1}{16\pi^{2}}\frac{1}{(y+1)^{3}}\int^{\omega_{c1}}_{2m_{s}}d\nu_{+}e^{-\frac{\nu_{+}}{T}}
[\nu^{4}_{+} + 2 m_{s}(y+1)\nu^{3}_{+}\nonumber\\&+ 3 m^{2}_{s}(y+1)\nu^{2}_{+}]
+\frac{\langle g_{s} \bar{s}\sigma\cdot G s\rangle}{12}(1-\frac{m_{s}}{8T})\nonumber\\&
-\frac{T}{48}\frac{y-5}{(y+1)^{2}}\langle\frac{\alpha_{s}}{\pi}GG\rangle.
\end{align}

\section{Numerical results and discussions}\label{sec4}
Now let us evaluate the sum rules derived in the previous section numerically. First, we specify the input parameters into our calculation. For the vacuum condensation parameters, we adopt the standard values: $\label{qcond}\langle\overline{q}q\rangle=-(0.24)^{3}\,\mbox{GeV}^{3}$, $\label{gcond}\langle\alpha_{s}GG\rangle=0.04\,\mbox{GeV}^{4}$, and $\label{scond}\langle\bar{s}s\rangle=(0.8\pm 0.2)\,\langle\overline{q}q\rangle$. The mass of the strange quark is $m_{s}=150\,\mbox{MeV}$. For masses of the initial $B_{s1}$, $B^{*}_{s2}$, $B_{s0}$, and $B'_{s1}$ mesons, we use $M_{B_{s1}}=5829.4\,\mbox{MeV}$,  $M_{B^{*}_{s2}}=5839.7\,\mbox{MeV}$, $M_{B_{s0}}=5718\,\mbox{MeV}$ \cite{BEH03}, and $M_{B'_{s1}}=5765\,\mbox{MeV}$. For masses of the final $D_{s1}$ and $D^{*}_{s2}$ mesons, we use $M_{D_{s1}}=2535.4\,\mbox{MeV}$ and $M_{D^{*}_{s2}}=2572.6\,\mbox{MeV}$ \cite{PDG10}.

\begin{figure}
\begin{center}
\begin{tabular}{ccc}
\begin{minipage}{7cm} \epsfxsize=7cm
\centerline{\epsffile{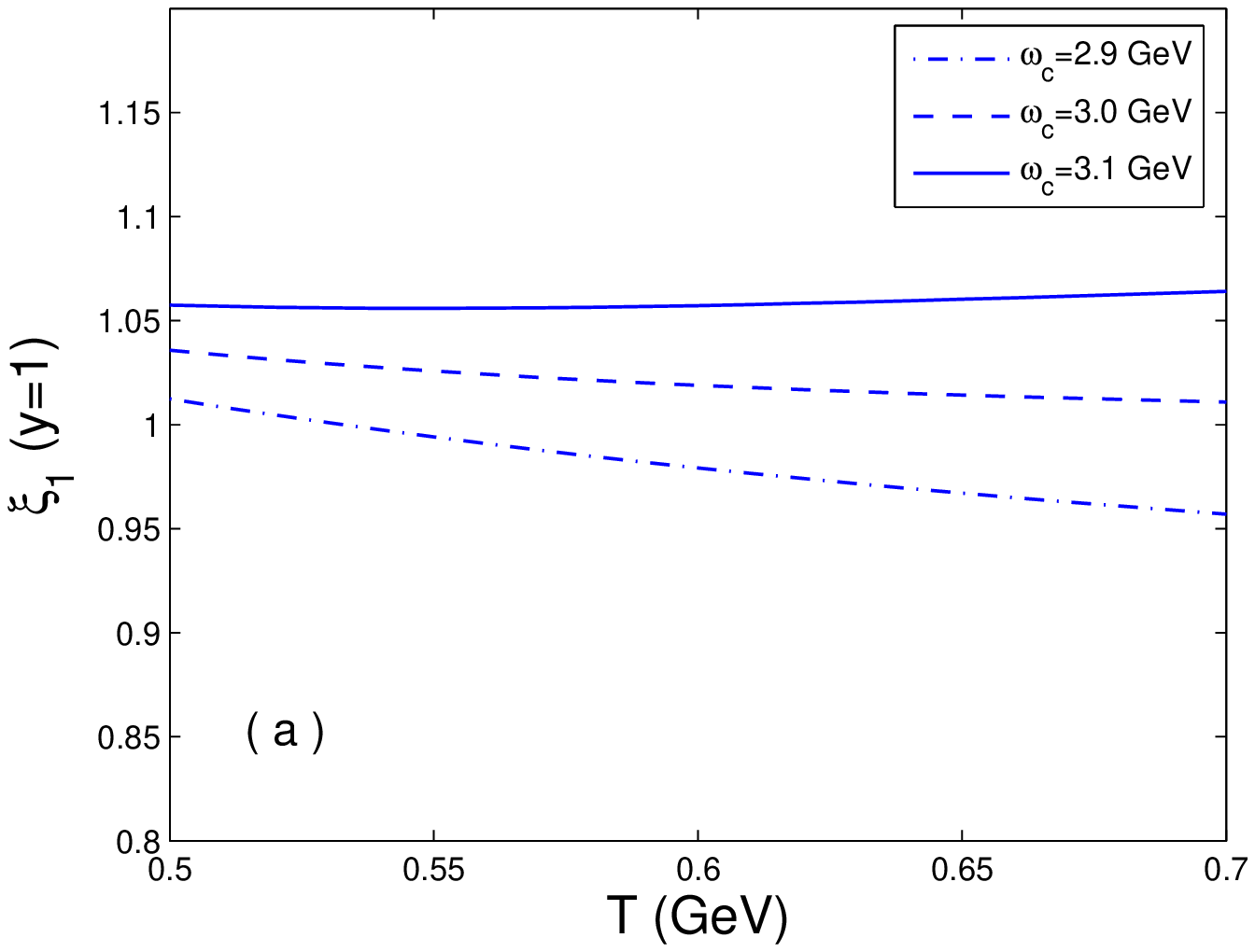}}
\end{minipage}& &
\begin{minipage}{7cm} \epsfxsize=7cm
\centerline{\epsffile{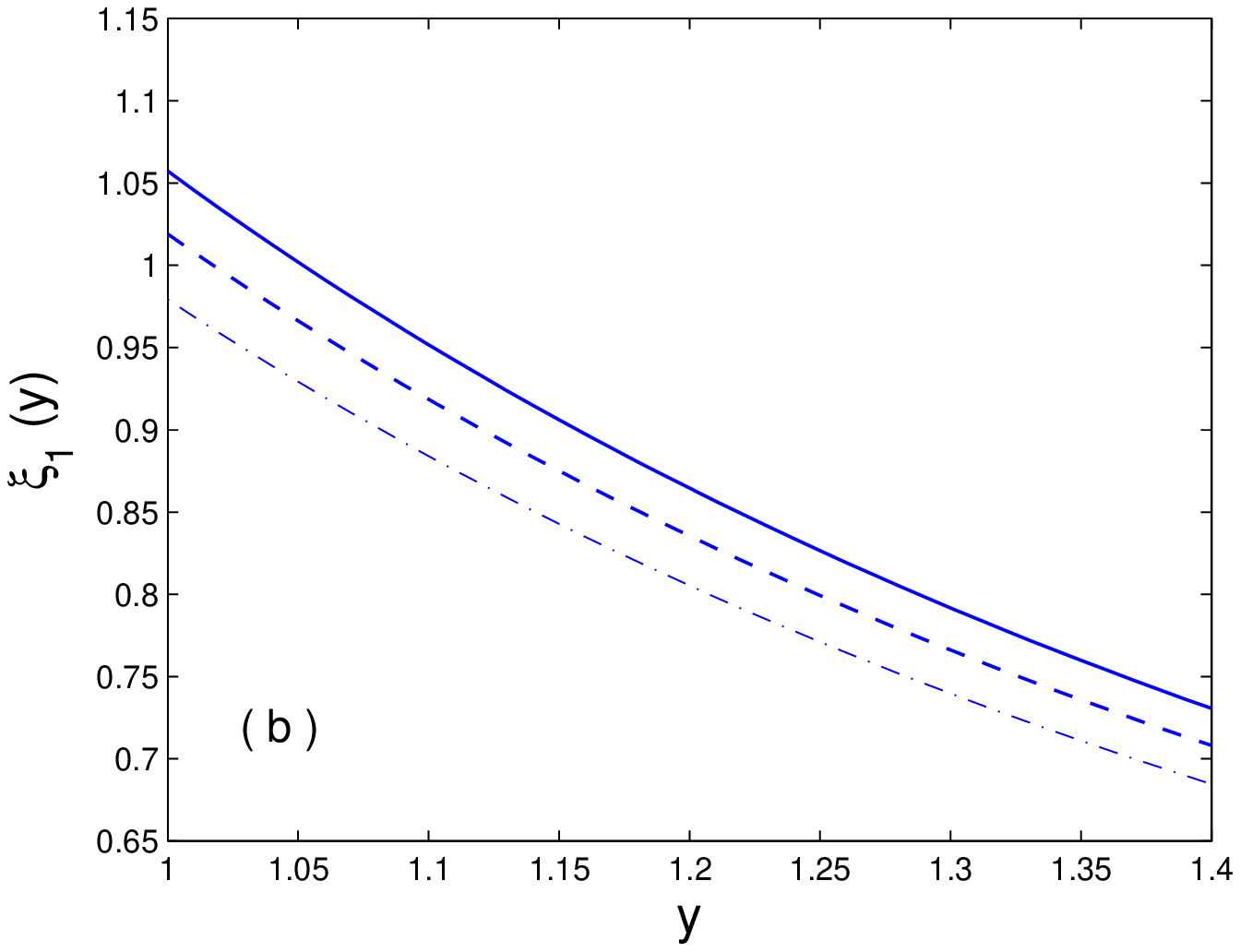}}
\end{minipage}
\end{tabular}
\caption{(a) {\it Dependence of  $\xi_{1}(y)$  on Borel parameter $T$ at $y=1$.} (b) {\it Prediction for the Isgur-Wise functions $\xi_{1}(y)$ at} $T=0.6\,\mbox{GeV}$.}\label{fig1}
\end{center}
\end{figure}

\begin{figure}
\begin{center}
\begin{tabular}{ccc}
\begin{minipage}{7cm} \epsfxsize=7cm
\centerline{\epsffile{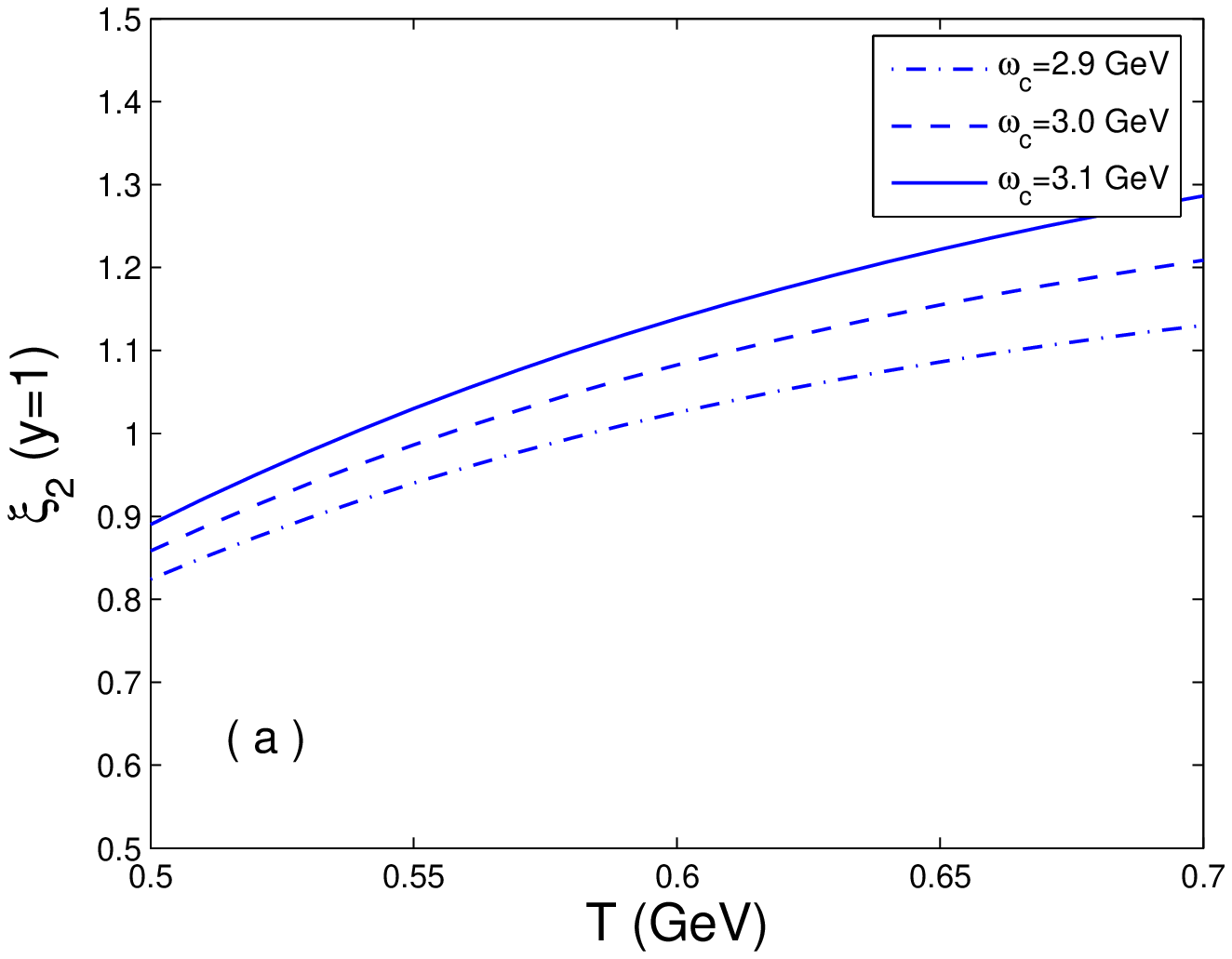}}
\end{minipage}& &
\begin{minipage}{7cm} \epsfxsize=7cm
\centerline{\epsffile{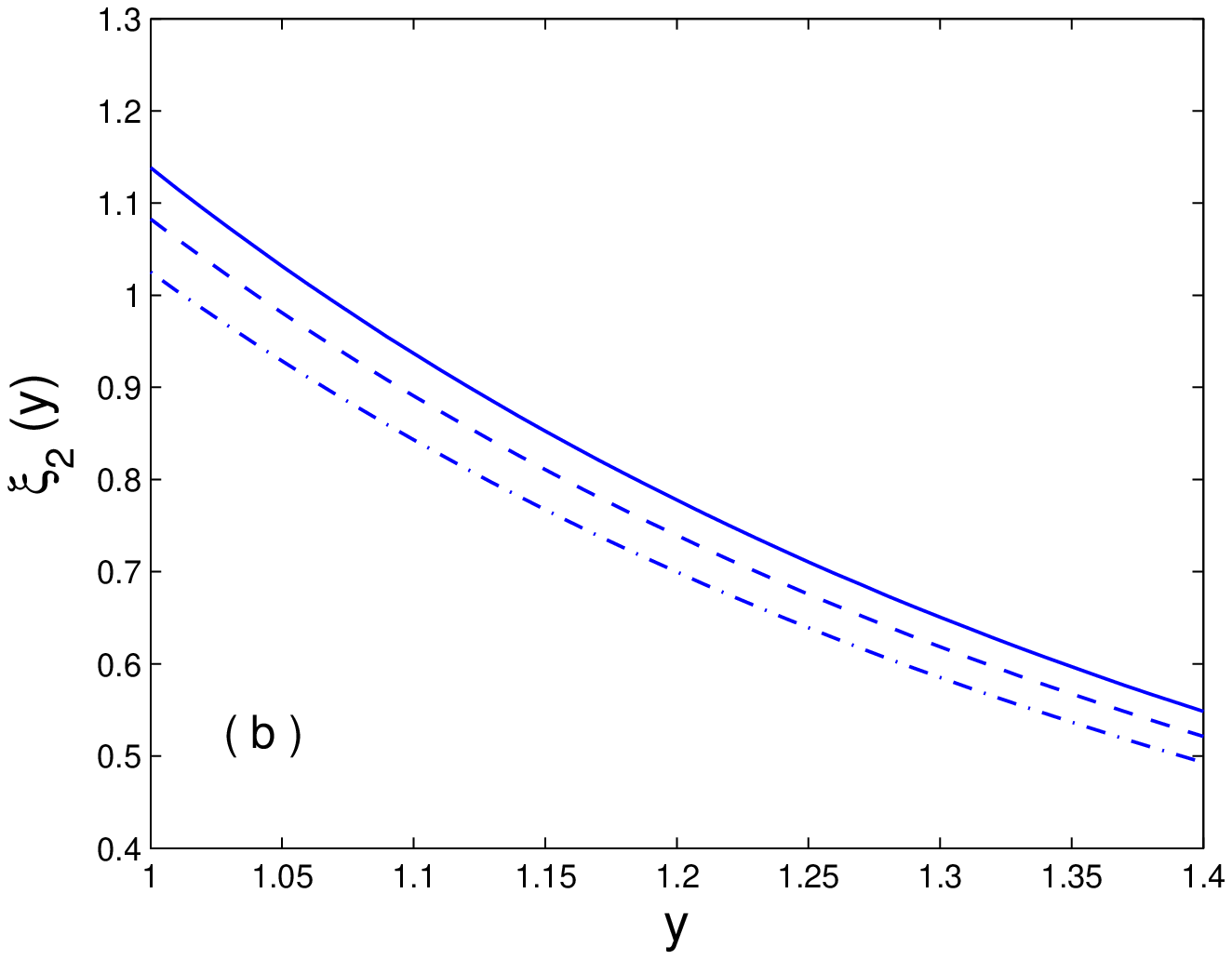}}
\end{minipage}
\end{tabular}
\caption{(a) {\it Dependence of $\xi_{2}(y)$ on Borel parameter $T$ at $y=1$.} (b) {\it Prediction for the Isgur-Wise functions $\xi_{2}(y)$ at} $T=0.6\,\mbox{GeV}$.}\label{fig2}
\end{center}
\end{figure}

\begin{figure}
\begin{center}
\begin{tabular}{ccc}
\begin{minipage}{7cm} \epsfxsize=7cm
\centerline{\epsffile{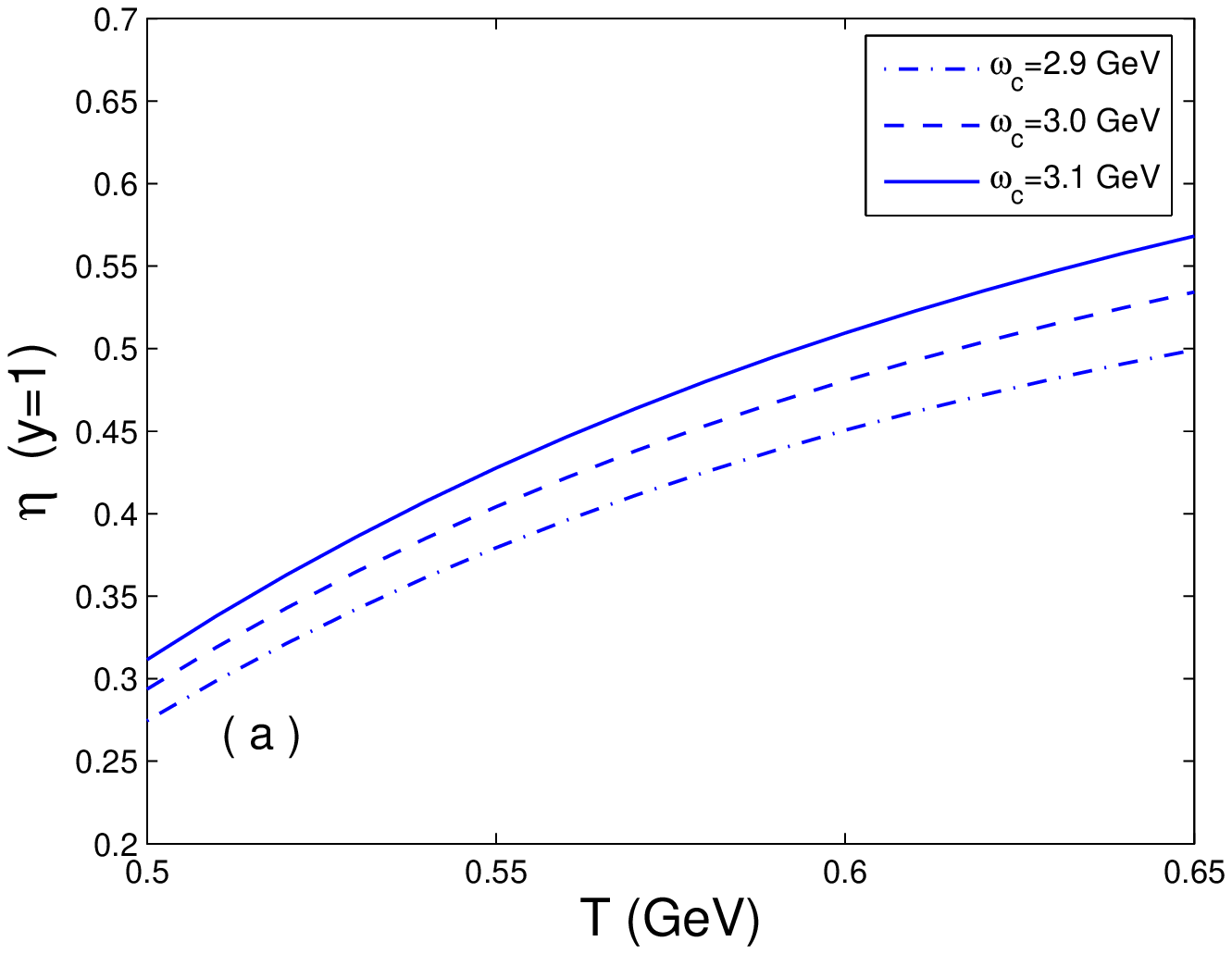}}
\end{minipage}& &
\begin{minipage}{7cm} \epsfxsize=7cm
\centerline{\epsffile{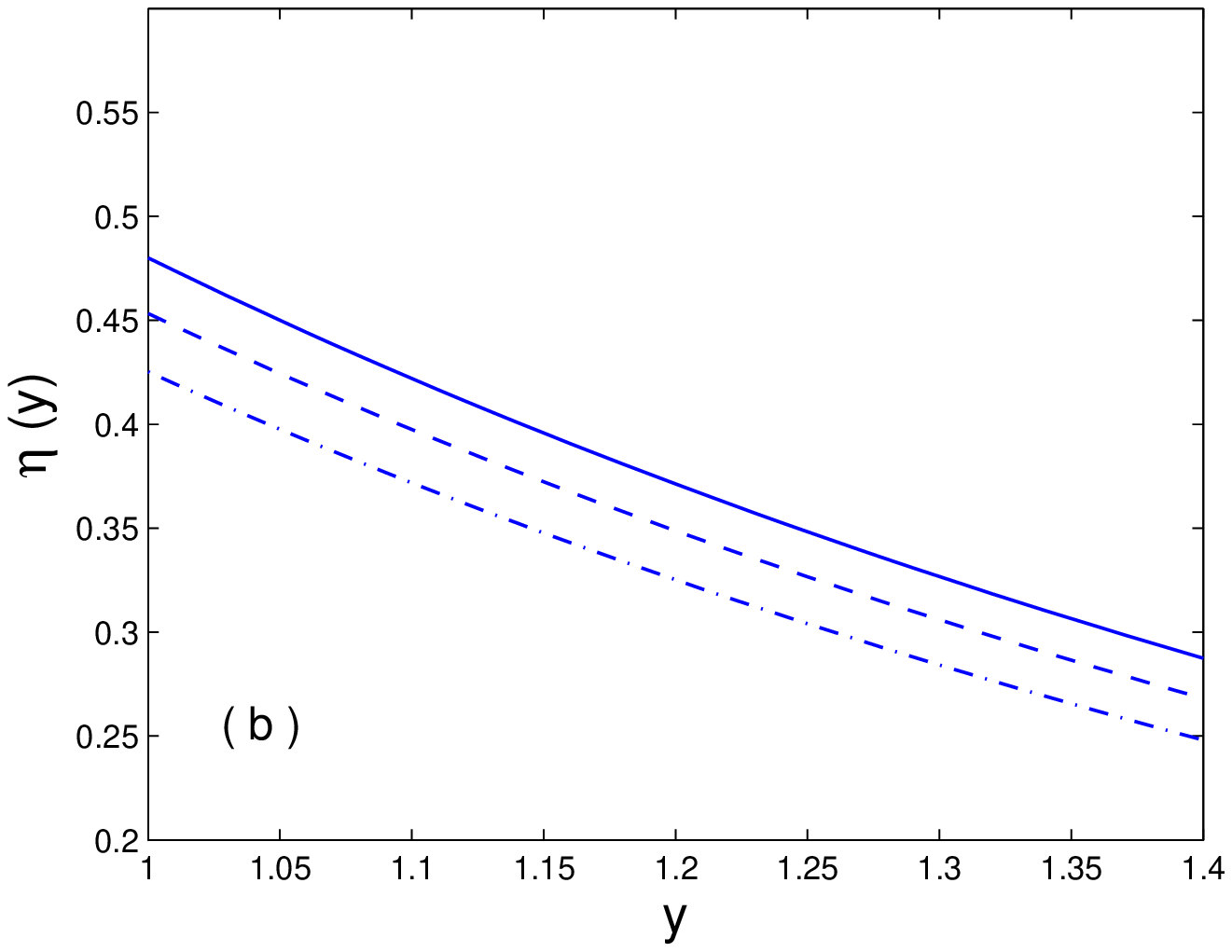}}
\end{minipage}
\end{tabular}
\caption{(a) {\it Dependence of $\eta(y)$ on Borel parameter $T$ at $y=1$.} (b) {\it Prediction for the Isgur-Wise functions $\eta(y)$ at} $T=0.58\,\mbox{GeV}$.}\label{fig3}
\end{center}
\end{figure}

In order to obtain information of Isgur-Wise function $\xi_{1,2}(y)$ and $\eta(y)$ with less systematic uncertainty, we can divide the three-point sum rules (\ref{rule1}), (\ref{rule2}), and (\ref{rule3}) with the square roots of relevant two-point sum rules for the decay constants, as many authors did \cite{Neu92,Hua99,Col00}. This can-not only reduce the number of input parameters but also improve stabilities of the three-point sum rules. The two-point QCD sum rule we need here are \cite{DHLZ03}
\begin{align}\label{conrule1}
f^{2}_{1,+,3/2}e^{-2\bar{\Lambda}_{+,3/2}/T}= & \frac{1}{64\pi^{2}}\int^{\omega_{c}}_{2m_{s}}d\nu e^{-\frac{\nu}{T}} (\nu^{4}+2m_{s}\nu^{3})-\frac{1}{12}m^{2}_{0}\langle\bar{s}s\rangle-\frac{1}{32}\langle\frac{\alpha_{s}}{\pi}GG\rangle T.
\end{align}
and
\begin{align}\label{conrule2}
f^{2}_{0,+,1/2}e^{-2\bar{\Lambda}_{0,+,1/2}/T}=& \frac{3}{64\pi^{2}}\int^{\omega_{c0}}_{2m_{s}}d\nu e^{-\frac{\nu}{T}} (\nu^{4}+2m_{s}\nu^{3}-6m^{2}_{s}\nu^{2}-12m^{3}_{s}\nu) -\frac{1}{16}m^{2}_{0}\langle \bar{s}s\rangle \nonumber\\& \times(1-\frac{m_{s}}{T}+\frac{4}{3}\frac{m^{2}_{s}}{T^{2}})+\frac{3}{8}m^{2}_{s}\langle \bar{s}s\rangle-\frac{m_{s}}{16}\langle\frac{\alpha_{s}}{\pi}GG\rangle.
\end{align}
After the divisions have been done, the Isgur-Wise functions $\xi_{1}(y)$, $\xi_{2}(y)$, and $\eta(y)$ depend only on the Borel parameter $T$ and the continuum thresholds. The determination of the Borel parameter is an important step of sum rules. After a careful analysis, we find the sum rules for $\xi_{1}(y)$ and $\xi_{2}(y)$ have a common sum rule ``window": $0.55\,\mbox{GeV} < T <0.65\,\mbox{GeV}$, which overlaps with that of the two-point sum rule (\ref{conrule1}) \cite{DHLZ03}. For the sum rule of $\eta(y)$, we choose the ``window" as $0.5\,\mbox{GeV} < T < 0.65\,\mbox{GeV}$. Note that the Borel parameter in the three-point sum rules is twice that in the two-point sum rules. In the evaluation, we have taken $2.9 \, \mbox{GeV} < \omega_{c} < 3.1 \, \mbox{GeV}$ and $2.9 \, \mbox{GeV} < \omega_{c0} < 3.1 \, \mbox{GeV}$. The regions of these continuum thresholds are fixed by analyzing the corresponding two-point sum rules \cite{DHLZ03}. Following the discussions in Refs. \cite{Blo93,Neu92}, the upper limit $\omega_{c}$ for $\nu_{+}$ in (\ref{rule1}) and (\ref{rule2}) is just the same as that for $\nu$ in (\ref{conrule1}). For $\omega_{c1}$ in Eq. (\ref{rule3}), it should be in the region $\frac{1}{2}[(y+1)-\sqrt{y^{2}-1}]\omega_{c0}\leqslant\omega_{c1}\leqslant\frac{1}{2} (\omega_{c0} +\omega_{c})$. So it can also be fixed in the region $2.9\,\mbox{GeV}<\omega_{c1}<3.1\,\mbox{GeV}$. The results are shown in Fig. \ref{fig1}, Fig. \ref{fig2}, and Fig. \ref{fig3}, where we have fixed $\omega_{c}=3.0\,\mbox{GeV}$ in the two-point sum rule (\ref{conrule1}) and $\omega_{c0}=3.0\,\mbox{GeV}$ in Eq. (\ref{conrule2}).

The curves for $\xi_{1}(y)$, $\xi_{2}(y)$, and $\eta(y)$ shown in the figures above can be parametrized by the linear approximations
\begin{align}
\label{linear1}
\xi_{1}(y)&=\xi_{1}(1)[1-\rho^{2}_{\xi_{1}}(y-1)],\text{ }
\xi_{1}(1)=1.03\pm0.03,\text{ }\rho^{2}_{\xi_{1}}=0.75\pm0.04;
\\ \label{linear2}
\xi_{2}(y)&=\xi_{2}(1)[1-\rho^{2}_{\xi_{2}}(y-1)],\text{ }
\xi_{2}(1)=1.00\pm0.03,\text{ }\rho^{2}_{\xi_{2}}=1.4\pm0.1;
\\ \label{linear3}
\eta(y)&=\eta(1)[1-\rho^{2}_{\eta}(y-1)],\text{ }
\eta(1)=0.45\pm0.03,\text{ }\rho^{2}_{\eta}=1.02\pm0.03.
\end{align}
The errors mainly come from the uncertainty due to $\omega_{c}$'s and $T$. It is difficult to estimate the systematic errors which are brought in by the quark-hadron duality. Using the linear approximates for the universal form factors above, we can calculate the semileptonic decay rates of processes $B_{s1}(B^{*}_{s2})\rightarrow D_{s1}(D^{*}_{s2})\ell\overline{\nu}$ and $B_{s0}(B'_{s1})\rightarrow D_{s1}(D^{*}_{s2})\ell\overline{\nu}$. For this purpose, we have to derive firstly the formulas for the differential decay rates of these processes in terms of the Isgur-Wise functions $\xi_{1}(y)$, $\xi_{2}(y)$, and $\eta(y)$ from the matrix elements (\ref{matrix9})-(\ref{matrix8}) given in Sec. \ref{sec2}. After some derivation, the formulas of the differential decay rates of the processes $B_{s1}(B^{*}_{s2})\rightarrow D_{s1}(D^{*}_{s2})\ell\overline{\nu}$ appear as
\begin{align}
\label{rate1}
\frac{d\Gamma}{dy}(B_{s1}\rightarrow D_{s1}\ell\overline{\nu})= &
\frac{G^{2}_{F}|V_{cb}|^{2}m^{2}_{B_{s1}}m^{3}_{D_{s1}}}{3^{4}\times2^{5}\pi^{3}}\{\xi_{1}(y)^{2}
[(r_{1}^{2}+1)(26y^{3}+18y^{2}+21y-11)-2r_{1}\nonumber\\&\times(16y^{4}+8y^{3}-12y^{2}-y+43)]+2\xi_{2}(y)^{2} (y^{2}-1)^{2}[(r_{1}^{2}+1)(13y-4)\nonumber\\&-2r_{1}(8y^{2}-4y+5)]-2\xi_{1}(y)\xi_{2}(y)
(y^{2}-1)[(r_{1}^{2}+1)(26y^{2}+5y-13)\nonumber\\&-2r_{1}(16y^{3}-3y+5)]\},
\\\label{rate2}
\frac{d\Gamma}{dy}(B_{s1}\rightarrow D^{*}_{s2}\ell\overline{\nu})= &
\frac{G^{2}_{F}|V_{cb}|^{2}m^{2}_{B_{s1}}m^{3}_{D^{*}_{s2}}}{3^{4}\times2^{5}\pi^{3}}\{\xi_{1}(y)^{2}
[(r_{2}^{2}+1)(46y^{3}+54y^{2}+159y+11)-2r_{2}\nonumber\\&\times(32y^{4}+40y^{3}+156y^{2}+25y+17)] +2\xi_{2}(y)^{2}(y^{2}-1)^{2}[(r_{2}^{2}+1)(23y \nonumber\\&+4)-2r_{2}(16y^{2}+4y+7)]
-2\xi_{1}(y)\xi_{2}(y)(y^{2}-1)[(r_{2}^{2}+1)(46y^{2}+31y \nonumber\\&+13)-2r_{2}(32y^{3}+24y^{2}+27y+7)]\},
\\\label{rate3}
\frac{d\Gamma}{dy}(B^{*}_{s2}\rightarrow D_{s1}\ell\overline{\nu})= &
\frac{G^{2}_{F}|V_{cb}|^{2}m^{2}_{B^{*}_{s2}}m^{3}_{D_{s1}}}{5\times3^{3}\times2^{5}\pi^{3}}\{\xi_{1}(y)^{2}
[(r_{3}^{2}+1)(46y^{3}+54y^{2}+159y+11)-2r_{3} \nonumber\\&\times(32y^{4}+40y^{3}+156y^{2}+25y+17)] +2\xi_{2}(y)^{2}(y^{2}-1)^{2}[(r_{3}^{2}+1)(23y \nonumber\\&+4)-2r_{3}(16y^{2}+4y+7)]
-2\xi_{1}(y)\xi_{2}(y)(y^{2}-1)[(r_{3}^{2}+1)(46y^{2}+31y \nonumber\\&+13)-2r_{3}(32y^{3}+24y^{2}+27y
+7)]\},
\\\label{rate4}
\frac{d\Gamma}{dy}(B^{*}_{s2}\rightarrow D^{*}_{s2}\ell\overline{\nu})= &
\frac{G^{2}_{F}|V_{cb}|^{2}m^{2}_{B^{*}_{s2}}m^{3}_{D^{*}_{s2}}}{5\times3^{3}\times2^{5}\pi^{3}}\{\xi_{1}(y)^{2}
[(r_{4}^{2}+1)(74y^{3}+66y^{2}+141y-11)-2r_{4} \nonumber\\&\times(48y^{4}+40y^{3}+84y^{2}+15y+83)] +2\xi_{2}(y)^{2} (y^{2}-1)^{2}[(r_{4}^{2}+1)(37y \nonumber\\&-4)+r_{4}(-48y^{2}+8y-26)]-2\xi_{1}(y) \xi_{2}(y) (y^{2}-1)[(r_{4}^{2}+1) (74y^{2} \nonumber\\&+29y-13)-r_{4}(96y^{3}+32y^{2}+26y+26)]\},
\end{align}
while for the processes $B_{s0}(B'_{s1})\rightarrow D_{s1}(D^{*}_{s2})\ell\overline{\nu}$, they can be found to be
\begin{align}
\label{rate5}
\frac{d\Gamma}{dy}(B_{s0}\rightarrow D_{s1}\ell\overline{\nu})= &
\frac{G^{2}_{F}|V_{cb}|^{2}m^{2}_{B_{s0}}m^{3}_{D_{s1}}}{3^{2}\times2^{3}\pi^{3}}|\eta(y)|^2(y-1)^{5/2}(y+1)^{3/2} [(r_{5}^{2}+1)(2y+1)\nonumber\\&-2r_{5}(y^2+y+1)],
\\\label{rate6}
\frac{d\Gamma}{dy}(B_{s0}\rightarrow D_{s2}\ell\overline{\nu})= &
\frac{G^{2}_{F}|V_{cb}|^{2}m^{2}_{B_{s0}}m^{3}_{D_{s2}}}{3^{2}\times2^{3}\pi^{3}}|\eta(y)|^2(y-1)^{5/2}(y+1)^{3/2} [(r_{6}^{2}+1)(4y-1)\nonumber\\&-2r_{6}(3y^2-y+1)],
\\\label{rate7}
\frac{d\Gamma}{dy}(B'_{s1}\rightarrow D_{s1}\ell\overline{\nu})= &
\frac{G^{2}_{F}|V_{cb}|^{2}m^{2}_{B'_{s1}}m^{3}_{D_{s1}}}{3^{3}\times2^{3}\pi^{3}}|\eta(y)|^2(y-1)^{5/2}(y+1)^{3/2} [(r_{7}^{2}+1)(7y-1)\nonumber\\&-2r_{7}(5y^2-y+2)],
\\\label{rate8}
\frac{d\Gamma}{dy}(B'_{s1}\rightarrow D_{s2}\ell\overline{\nu})= &
\frac{G^{2}_{F}|V_{cb}|^{2}m^{2}_{B'_{s1}}m^{3}_{D_{s2}}}{3^{3}\times2^{3}\pi^{3}}|\eta(y)|^2(y-1)^{5/2}(y+1)^{3/2} [(r_{8}^{2}+1)(11y+1)\nonumber\\&-2r_{8}(7y^2+y+4)],
\end{align}
where $r_{i}$ ($i=1, \cdots, 8$) is the ratio between the mass of the final $D_{s}$ meson and that of the initial $B_{s}$ meson in each process, e.g., $r_{1}=\frac{M_{D_{s1}}}{M_{B_{s1}}}$. The maximal values of $y$ for these semileptonic processes are given in Table \ref{table1}.
\begin{table}[h]
\caption{The maximal value of $y$ for each process: $y_{\text{max}}=(1+r_{i}^{2})/2r_{i}$ ($i=1, 2, \cdots, 8$).}
\begin{center}
\begin{tabular}{ccccccccc}
\hline \hline
 & & $B_{s1}$ & & $B_{s2}$ & & $B_{s0}$ & & $B'_{s1}$\\
\hline
$D_{s1}\ell\overline{\nu}$  & & 1.36707 & & 1.36872 & & 1.34934 & & 1.3568 \\
$D_{s2}\ell\overline{\nu}$ & & 1.35364 & & 1.35525 & & 1.33628 & & 1.34358 \\
\hline \hline
\end{tabular}
\end{center}\label{table1}
\end{table}
In addition, we need $|V_{cb}|=0.04$ and $G_{F}=1.166\times10^{-5}\mbox{GeV}^{-2}$. By integrating the differential decay rates over the kinematic region $1.0 \leq y \leq y_{\text{max}}$, we get the decay widths of these semileptonic decay modes which are listed in Table \ref{table2}.
\begin{table}[h]
\caption{Predictions for the decay widths and branching ratios }
\begin{center}
\begin{tabular}{ccccc}
  \hline
  \hline
  Decay mode & & Decay width (GeV)  & & Branching ratio  \\
  \hline
  $B_{s1}\rightarrow D_{s1}\ell\overline{\nu}$ & & $(2.1\pm 0.2)\times10^{-14}$ & & $\sim 10^{-10}$ \\
  $B_{s1}\rightarrow D^{*}_{s2}\ell\overline{\nu}$ & & $(2.8\pm 0.2)\times10^{-14}$ & & $\sim 10^{-10}$ \\
  $B^{*}_{s2}\rightarrow D_{s1}\ell\overline{\nu}$ & & $(1.8\pm 0.1)\times10^{-14}$ & & $\sim 10^{-11}$ \\
  $B^{*}_{s2}\rightarrow D^{*}_{s2}\ell\overline{\nu}$ & & $(2.9\pm 0.2)\times10^{-14}$ & & $\sim 10^{-11}$ \\
  \hline
  $B_{s0}\rightarrow D_{s1}\ell\overline{\nu}$ & & $(1.1\pm 0.2)\times10^{-16}$ & & $\sim 10^{-12}$ \\
  $B_{s0}\rightarrow D^{*}_{s2}\ell\overline{\nu}$ & & $(1.0\pm 0.2)\times10^{-16}$ & & $\sim 10^{-12}$ \\
  $B'_{s1}\rightarrow D_{s1}\ell\overline{\nu}$ & & $(0.8\pm 0.1)\times10^{-16}$ & & $\sim 10^{-12}$ \\
  $B'_{s1}\rightarrow D^{*}_{s2}\ell\overline{\nu}$ & & $(1.5\pm 0.3)\times10^{-16}$ & & $\sim 10^{-12}$ \\
  \hline
  \hline
\end{tabular}
\end{center}\label{table2}
\end{table}
Although the widths of $B_{s1}$ and $B^{*}_{s2}$ have not yet been measured experimentally, they were estimated early in Ref. \cite{OPAL95} to be around 1 MeV. Theoretically, their strong decays were investigated in Ref. \cite{LCL09}. As we know, the main decay modes of these excited $B_{s}$ mesons are strong decays. Therefore we can approximately take the strong decay widths as the total widths for an estimation of order of the branching ratios of these processes. In fact, the two-body strong decay widths of $B_{s1}$ and $B^{*}_{s2}$ are computed to be $98\mbox{keV}$ and $5\mbox{MeV}$ in Ref. \cite{LCL09}. The masses of $B_{s0}$ and $B'_{s1}$ are considered to lie below the thresholds of $B^{*}K$ and $BK$, so their main decay modes are isospin violating decays and radiative decays and are supposed to have width of about $100$ keV \cite{BEH03}. Using these widths, we estimate the orders of the branching ratios of the semileptonic decays (see Table \ref{table2}). Note that the semileptonic decay branching ratios of $B_{s0}$ and $B'_{s1}$ should be 3 orders of magnitude lower if their masses lie above the thresholds of $B^{*}K$ and $BK$. As we can see from Table \ref{table2}, the semileptonic decay widths of ($B_{s1}$, $B^{*}_{s2}$) into ($D_{s1}$, $D^{*}_{s2}$) are significantly larger than those of ($B_{s0}$, $B'_{s1}$) into ($D_{s1}$, $D^{*}_{s2}$), as was expected. The present precision of the experimental measurement of the branching ratio of the $B_{s}$ mesons has reached up to $10^{-7} \sim 10^{-8}$ \cite{PDG10}. Therefore, all the branching ratios of these processes are so small that it is difficult to find them in experiments. However, we can still expect that with the precision of the detector improved, the decays $B_{s1}\rightarrow D_{s1}(D_{s2})\ell\overline{\nu}$ might be seen in the LHCb experiment.

In summary, we have performed a study of the semileptonic decays of the orbitally $P$-wave excited $B_{s}$ meson states $B^{**}_{s}$, including the newly observed $B_{s1}(5830)$ and $B^{*}_{s2}(5840)$ mesons into the $T$ doublet of $D^{**}_{s}$ mesons, $D_{s1}(2536)$ and $D^{*}_{s2}(2573)$, within the framework of HQET. We employ QCD sum rules to estimate the leading-order universal form factors describing these weak transitions. Different from the semileptonic processes of lower Heavy-light meson states, two universal Isgur-Wise functions $\xi_{1}(y)$ and $\xi_{2}(y)$ are need to parameterize the hadronic matrix elements in the weak transitions $B_{s1}(B^{*}_{s2})\rightarrow D_{s1}(D^{*}_{s2})\ell\overline{\nu}$ at the leading order of the heavy quark expansion. The predicted branching ratios of these processes are prohibitively tiny, so it is difficult to find them in experiments. The decay widths of $B_{s1}\rightarrow D_{s1}(D_{s2})\ell\overline{\nu}$ are comparatively large so that we expect they might be seen in the future LHCb experiment. It is worth noting that the Isgur-Wise functions $\xi_{1}(y)$ and $\xi_{2}(y)$ which parameterize the hadronic matrix elements of the weak currents between the states in the $T$ doublet of $\bar{b}s$ system and those in the same doublet of $\bar{c}s$ system approximately satisfy the normalization condition, $\xi_{1}(y=1)=1$ and $\xi_{2}(y=1)=1$, which is implied by the heavy quark flavor symmetry at the leading order of the heavy quark expansion.

\begin{acknowledgments}
This work was supported in part by the National Natural Science Foundation of China under Contract No. 10975184.
\end{acknowledgments}

\end{document}